# Oxygen Potential Transition in Mixed Conducting Oxide Electrolyte

Yanhao Dong and I-Wei Chen*

*Department of Materials Science and Engineering, University of Pennsylvania, Philadelphia, PA 19104, USA*

**Abstract**

It is generally assumed that oxygen potential in a thin oxide electrolyte follows a linear distribution between electrodes. Jacobsen and Mogensen have shown, however, that this is not the case for thin zirconia membranes in solid oxide electrochemical cells. Here we demonstrate that there is a ubiquitous oxygen potential transition rooted in the p-type/n-type transition of electronic conductivity inside mixed conducting oxides, and that the transition is extremely sensitive to electrode potential and current density. It is also remarkably sensitive to the conductivity ratio of electrons and holes, as well as their association with lattice oxygens and vacancies, which tends to increase the oxygen flow. Direct evidence of a sharp oxygen potential transition has been found in an equally sharp grain size transition in electrically loaded zirconia. More broadly speaking, the oxygen potential transition is akin to a first-order phase transition. Therefore, it will suffer interface instability, especially in high-current-density devices. These findings provide new opportunities to understand several disparate observations in the literature, from microstructural degradation and stress distribution in solid oxide fuel/electrolyzer cells, to field-assisted sintering, to conducting filaments in resistance memory, to dendrite formation in electrochemical cells.

*Corresponding Author Information**

**Tel: +1-215-898-5163; Fax: +1-215-573-2128**

***E-mail address*: iweichen@seas.upenn.edu (I-Wei Chen)**

***Postal address*:** Department of Materials Science and Engineering, University of Pennsylvania, LRSM Building, Room 424, 3231 Walnut St., Philadelphia, PA 19104-6272

## I. Introduction

Electrode overpotential and polarization is central to many electrochemical devices. For solid oxide fuel cells (SOFC) [1-6] and electrolyzer cells (SOEC) [5-7], it refers to a change in the local chemical potential of oxygen, or oxygen potential for short. Much attention has been paid to the discontinuous potential drop across the electrode/electrolyte interface (i.e. optimizing the electrode materials and microstructures), which drives anodic/cathodic half-reactions electrochemically and results in a voltage loss. However, the oxygen potential distribution inside the solid electrolyte is also important, because it affects the local chemistry and stoichiometry, hence critical to the chemical (e.g., degradation and decomposition [8,9]), microstructural (e.g., pore/bubble formation [10-16] and grain growth [15-17]) and mechanical (e.g., cracking and chemical expansion [18,19]) stability of the electrolyte and device. Overpotential at higher temperature than typically seen in SOFC/SOEC is also of interest, for example, to field assisted sintering, which uses a voltage bias to drive a large current across the sample or the mold as in spark plasma sintering [20,21] and flash sintering [22,23]. Since most ceramics are relatively poor thermal and electrical conductors, very high temperatures are often reached in field assisted sintering, meanwhile, electrode overpotentials are again required to convert electronic currents in the external circuit to ionic currents inside the electrolyte. For oxides, thus altered oxygen potentials inside the electrolyte will undoubtedly affect defect equilibrium and charge/mass transport, which in turn alters the microstructure. The solutions of oxygen potential distributions inside the solid electrolyte presented in the present study will help understand these phenomena.

Our model material is the prevailing solid electrolyte for oxygen transport, which is yttria-stabilized zirconia (YSZ) [1-7,10-12], specifically 8YSZ that contains 8 mol% of $Y_2O_3$. In YSZ, every two $Y^{3+}$ substitutions on the $Zr^{4+}$ sites are charge-compensated by one doubly charged oxygen vacancy. Under normal operating conditions of SOFC/SOEC, doubly charged oxygen vacancies are responsible for the ionic conductivity, which is independent of oxygen potential and >1,000 times higher than its electronic counterpart as shown in **Fig. 1** [24]. However, it is not uncommon for some parts of YSZ electrolyte to experience mixed ionic and electronic conduction under severe electrochemical reduction. If so, the oxygen potential distribution is expected to become non-linear for two reasons: (i) The dependence of electron/hole conductivity on oxygen potential in **Fig. 1** is non-linear [24]. (ii) An electron can associate with an oxygen vacancy, and likewise a hole with a lattice oxygen ion, thus altering their valences. In fact, there is a possibility that a part of the electronic conductivity in **Fig. 1** may actually arise from such associations. Nevertheless, for thin electrolytes under typical SOFC/SOEC operation conditions when electron and hole concentrations are low, Jacobsen and

Mogensen only considered (i) because (ii) is probably unimportant. [25]

More broadly, though, (ii) cannot be ignored in view of the following observations. Direct evidence for the existence of and conversion between $V_O^{\bullet\bullet}$, $V_O^{\bullet}$ and $V_O^{\times}$, known as color centers in reduced YSZ, was seen in electron spin resonance measurements, finding unpaired electrons. [26-29] These color centers—presumably F centers where one or more unpaired electrons are localized at the doubly charged oxygen vacancy $V_O^{\bullet\bullet}$—are apparently responsible for reduction-caused blacking widely observed in YSZ and other zirconia [30,31]. Formation of oxygen bubbles, which has been reported in several YSZ SOEC studies [10-13], also implies (ii) since each condensing lattice/interstitial oxygen ion must shed two electrons. Likewise, to form a vacancy void, which we recently observed in YSZ [14-16], each condensing oxygen vacancy must acquire two electrons. Lastly, although one would ordinarily expect migration of ions to be more difficult than migration of electrons or holes, $O^{2-}$ migration in YSZ actually has a lower activation energy (0.79 eV) than that of electrons (1.89 eV) and holes (1.05 eV) [24]. This unusual result could indicate that electrons and holes are somehow strongly trapped, presumably at vacancies/interstitials, forming complexes. Yet it has not been possible to examine whether these reactions will alter the spatial distribution of oxygen potential because direct experimental measurements of local oxygen potentials inside solid electrolytes are exceedingly difficult.

Having established a strong correlation between oxygen potential and grain growth kinetics in YSZ and related fluorite structure ceramics (reduction enhances grain growth), we have used grain size as an internal marker to map the oxygen potentials in YSZ electrolytes under various current densities, atmospheres, and electrode configurations. [14-17] As a result, we now have detailed information of oxygen potential distributions in an electrolyte for the first time.

Remarkably, it reveals a sharp oxygen potential transition—implicated by a sharp grain size distribution, one example shown in **Fig. 2**—typically at about half-way between the two electrodes (**Fig. 2** inset), and the transition becomes sharper as the current density increases or the electrode kinetics deteriorates. Qualitatively, the sharp transition can be understood from **Fig. 1**, which has a minimum in the combined red-blue curve (the combined electron-hole conductivity) at an intermediate oxygen potential: If the oxygen potential forces a steady-state electronic current and the current must go through the conductivity minimum, then most of the potential gradient must be spent at the conductivity minimum—this corresponds to a sharp oxygen potential transition.

Such a transition was indeed seen in the Jacobsen-Mogensen solution [25] and other literatures [32,33]. However, since these solutions did not consider internal reactions, they may not apply to our experimental conditions that experienced more severe electrochemical reductions and copious cavitation, with considerably larger current densities and thicker electrolytes. Therefore, we will present here a more complete solution that allows both (i) and (ii), with fully and partially ionized species, for a wide range of electrolyte thicknesses and current densities. More broadly, we will argue that the oxygen potential transition is akin to a first-order phase transition, and as such the narrow transition zone may be susceptible to diffusion-limited "interface" instability (e.g., dendritic instability in solidification [34]), which is fundamental for understanding such disparate phenomena as filament growth and breakdown of mixed conducting oxide electrolytes in devices from electrochemical cells, to ceramic capacitors [35], to memristors [36], over a wide range of temperatures. The experimental observations of such instabilities will be presented and analyzed in a separate paper using the idea developed here.

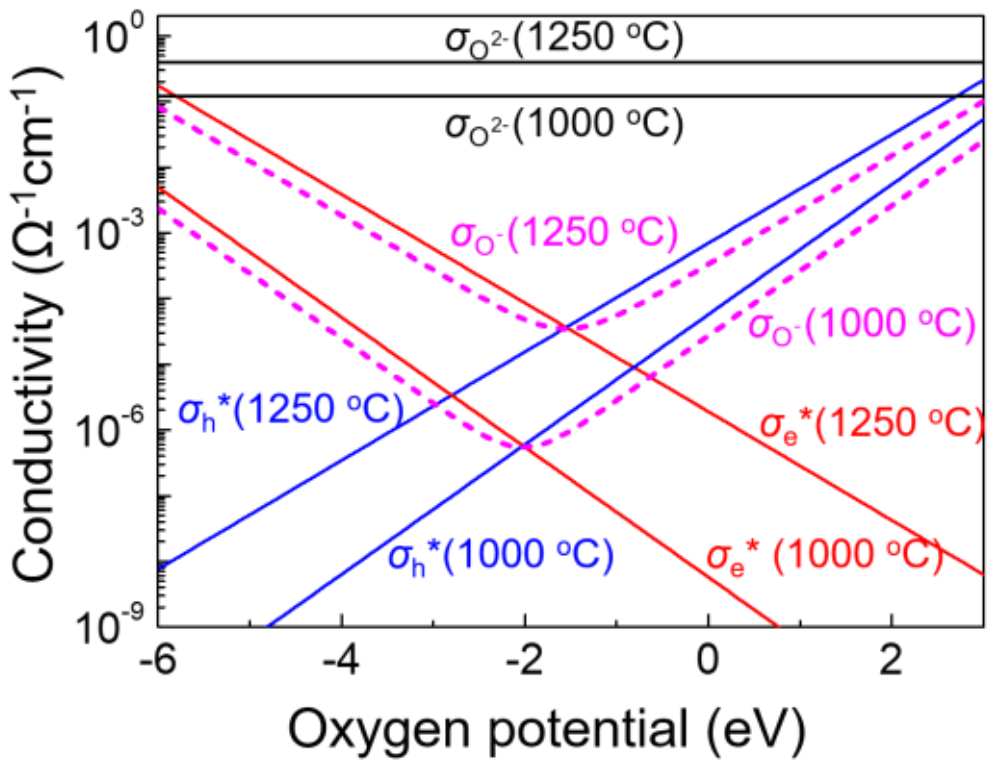

**Figure 1** Extrapolated conductivity data (taken directly from Ref. [24] without subtracting $\sigma_{O^-}$) for fully ionized oxygen ions ($\sigma_{O^{2-}}$ in black), electrons ($\sigma_e$* in red) and holes ($\sigma_h$* in blue) at two temperatures (1000 °C and 1250 °C). Computed conductivity for partially ionized oxygen ions ($\sigma_{O^-}$ in purple) are also shown for the case of $\alpha_e = \alpha_h = 0.475$ (see Section V for details). Oxygen potential is set to be 0 eV at 1 atm oxygen partial pressure.

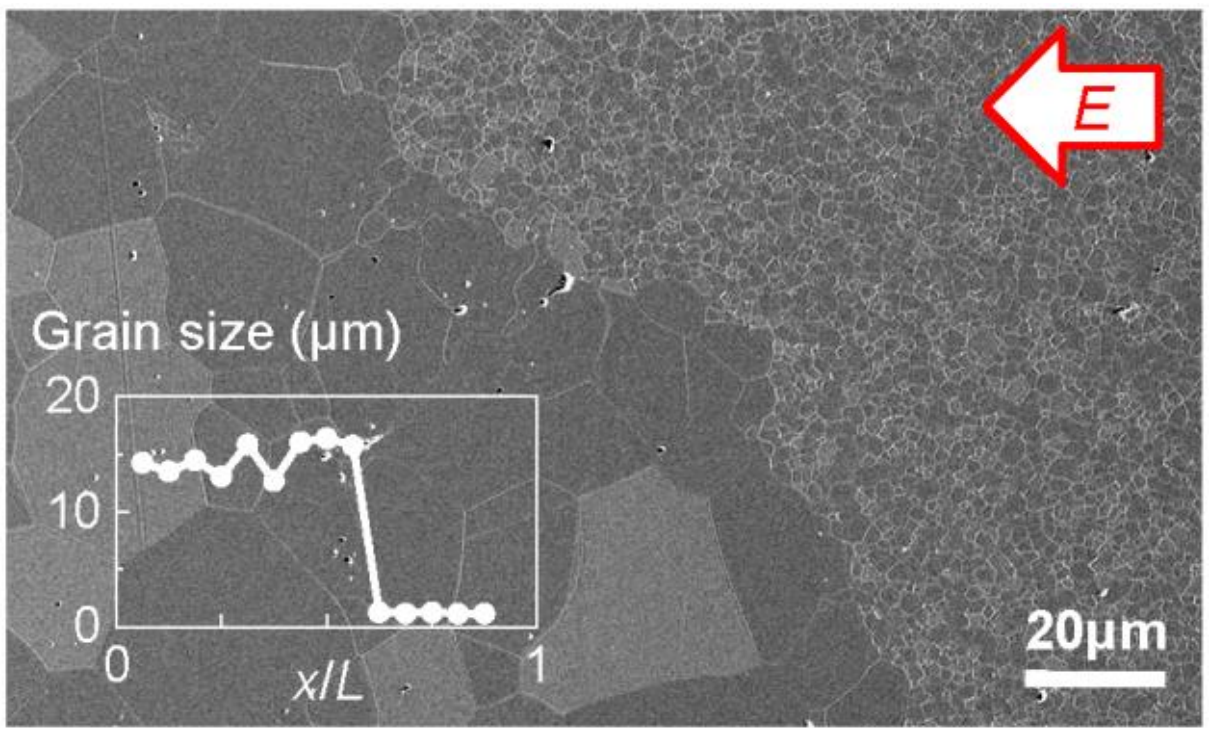

**Figure 2** Microstructure of 8YSZ after 10 h electrical testing showing a grain size transition halfway across the thickness. Inset: Grain size distribution along distance *x* from left electrode (cathode). Sample temperature: 1360 °C, thickness: $L$=1.72 mm, current density: −50 A/cm$^2$. For more details, see Ref. [14-17].

## II. Formulation of the problem

We treat YSZ as a prototypical oxygen conductor with mixed ionic and electronic conduction. In addition to the standard ion-diffusion mechanism in which $O^{2-}$ exchanges its location with a doubly-charged oxygen vacancy neighbor, $V_O^{\bullet\bullet}$, we shall allow the following possibilities. First, if an electron (e) is localized at the oxygen vacancy (equivalent to a singly-charged oxygen vacancy $V_O^{\bullet}$), then as the electron-tagged vacancy exchanges with a neighboring lattice oxygen ion $O^{2-}$, it amounts to the migration of $O^-$ instead of $O^{2-}$. Second, if a hole (h) is localized at a lattice $O^{2-}$ (equivalent to $O^-$), then the exchange of $O^-$ with an electron-free vacancy $V_O^{\bullet\bullet}$ again amounts to $O^-$ migration.

Note that in diffusion, an $O^-$ is not a hole, but is a sum of an $O^{2-}$ and a hole. This is because hole movement does not require atomic movement and cannot cause atomic diffusion, but $O^-$ movement must involve physical transport of oxygen and does cause atomic diffusion. Since chemical potential arises from mass conservation, and $O^-$ movement involves different species from those of hole movement, the chemical potential of an $O^-$ must differ from that of a hole. Therefore, in an oxygen ion conductor, one need to consider four charge carriers, $O^{2-}$, $O^-$, e and h, and five electrochemical potentials, $\tilde{\mu}_{O^{2-}}$, $\tilde{\mu}_{O^-}$, $\tilde{\mu}_e$, $\tilde{\mu}_h$ and the chemical potential of molecular oxygen, $\mu_{O_2}$. We shall limit ourselves to the one-dimensional problem along the *x* direction, for which the boundary condition at the electrode/electrolyte interfaces is $\mu'_{O_2}$ on the left and $\mu''_{O_2}$ on the right, assuming $\mu'_{O_2} < \mu''_{O_2}$. Transport in the electrodes and at the electrode/atmosphere interface is not considered.

We assume local equilibrium for the following reactions

$$O^{2-} = \frac{1}{2}O_2 + 2e \qquad (1)$$

$$e + h = \text{nil} \qquad (2)$$

$$O^{2-} = O^- + e \qquad (3)$$

They relate the five potentials with

$$\tilde{\mu}_{\mathrm{O^{2-}}} = \frac{1}{2}\mu_{\mathrm{O_2}} + 2\tilde{\mu}_{\mathrm{e}} \qquad (4)$$

$$\tilde{\mu}_{\mathrm{e}} + \tilde{\mu}_{\mathrm{h}} = 0 \qquad (5)$$

$$\tilde{\mu}_{\mathrm{O^{2-}}} = \tilde{\mu}_{\mathrm{O^{-}}} + \tilde{\mu}_{\mathrm{e}} \qquad (6)$$

leaving two independent potentials, or equivalently, two independent fluxes. Since molecular oxygen is not allowed to leak through the electrolyte, the steady-state charge flux, $j_{\mathrm{charge}}$, and chemical flux, $J_{\mathrm{O}}$, must be constant, with their ratio fixed. Therefore, at the steady state there is only one independent variable, either a potential such as $\mu_{\mathrm{O_2}}$, or a flux such as $j_{\mathrm{charge}}$ or $J_{\mathrm{O}}$.

To proceed, we express *charge* fluxes by all species

$$j_{\mathrm{O^{2-}}} = \frac{\sigma_{\mathrm{O^{2-}}}}{2e}\frac{d\tilde{\mu}_{\mathrm{O^{2-}}}}{dx} \qquad (7)$$

$$j_{\mathrm{e}} = \frac{\sigma_{\mathrm{e}}}{e}\frac{d\tilde{\mu}_{\mathrm{e}}}{dx} \qquad (8)$$

$$j_{\mathrm{h}} = -\frac{\sigma_{\mathrm{h}}}{e}\frac{d\tilde{\mu}_{\mathrm{h}}}{dx} = \frac{\sigma_{\mathrm{h}}}{e}\frac{d\tilde{\mu}_{\mathrm{e}}}{dx} \qquad (9)$$

$$j_{\mathrm{O^{-}}} = \frac{\sigma_{\mathrm{O^{-}}}}{e}\frac{d\tilde{\mu}_{\mathrm{O^{-}}}}{dx} = \frac{\sigma_{\mathrm{O^{-}}}}{e}\left(\frac{d\tilde{\mu}_{\mathrm{O^{2-}}}}{dx} - \frac{d\tilde{\mu}_{\mathrm{e}}}{dx}\right) \qquad (10)$$

In the above, $\sigma_i$ denotes the conductivity of species $i$. From which, we obtain

$$\begin{aligned} J_{\mathrm{O}} &= J_{\mathrm{O^{2-}}} + J_{\mathrm{O^{-}}} = -\frac{1}{2e} j_{\mathrm{O^{2-}}} - \frac{1}{e} j_{\mathrm{O^{-}}} \\ &= -\frac{1}{4e^2}\left(\sigma_{\mathrm{O^{2-}}} + 4\sigma_{\mathrm{O^{-}}}\right)\frac{d\tilde{\mu}_{\mathrm{O^{2-}}}}{dx} + \frac{1}{e^2}\sigma_{\mathrm{O^{-}}}\frac{d\tilde{\mu}_{\mathrm{e}}}{dx} \end{aligned} \qquad (11)$$

$$\begin{aligned} j_{\mathrm{charge}} &= j_{\mathrm{O^{2-}}} + j_{\mathrm{e}} + j_{\mathrm{h}} + j_{\mathrm{O^{-}}} \\ &= \frac{1}{2e}\left(\sigma_{\mathrm{O^{2-}}} + 2\sigma_{\mathrm{O^{-}}}\right)\frac{d\tilde{\mu}_{\mathrm{O^{2-}}}}{dx} + \frac{1}{e}\left(\sigma_{\mathrm{e}} + \sigma_{\mathrm{h}} - \sigma_{\mathrm{O^{-}}}\right)\frac{d\tilde{\mu}_{\mathrm{e}}}{dx} \end{aligned} \qquad (12)$$

which may be expressed in the matrix form,

$$\begin{bmatrix} -e^2 J_\mathrm{O} \\ ej_\mathrm{charge} \end{bmatrix} = \begin{bmatrix} \dfrac{\sigma_{\mathrm{O}^{2-}}}{4} + \sigma_{\mathrm{O}^-} & -\sigma_{\mathrm{O}^-} \\ \dfrac{\sigma_{\mathrm{O}^{2-}}}{2} + \sigma_{\mathrm{O}^-} & \sigma_\mathrm{e} + \sigma_\mathrm{h} - \sigma_{\mathrm{O}^-} \end{bmatrix} \begin{bmatrix} \dfrac{d\tilde{\mu}_{\mathrm{O}^{2-}}}{dx} \\ \dfrac{d\tilde{\mu}_\mathrm{e}}{dx} \end{bmatrix} \tag{13}$$

then inverted into

$$\begin{bmatrix} \dfrac{d\tilde{\mu}_{\mathrm{O}^{2-}}}{dx} \\ \dfrac{d\tilde{\mu}_\mathrm{e}}{dx} \end{bmatrix} = \frac{1}{\Delta} \begin{bmatrix} \sigma_\mathrm{e} + \sigma_\mathrm{h} - \sigma_{\mathrm{O}^-} & \sigma_{\mathrm{O}^-} \\ -\dfrac{\sigma_{\mathrm{O}^{2-}}}{2} - \sigma_{\mathrm{O}^-} & \dfrac{\sigma_{\mathrm{O}^{2-}}}{4} + \sigma_{\mathrm{O}^-} \end{bmatrix} \begin{bmatrix} -e^2 J_\mathrm{O} \\ ej_\mathrm{charge} \end{bmatrix} \tag{14}$$

Here the determinant $\Delta$ is

$$\Delta = \frac{\sigma_{\mathrm{O}^{2-}}}{4}\left(\sigma_\mathrm{e} + \sigma_\mathrm{h} + \sigma_{\mathrm{O}^-}\right) + \sigma_{\mathrm{O}^-}\left(\sigma_\mathrm{e} + \sigma_\mathrm{h}\right) \tag{15}$$

Using Eq. (4), we express $\dfrac{d\mu_{\mathrm{O}_2}}{dx}$ as

$$\begin{aligned} \frac{d\mu_{\mathrm{O}_2}}{dx} &= 2\frac{d\tilde{\mu}_{\mathrm{O}^{2-}}}{dx} - 4\frac{d\tilde{\mu}_\mathrm{e}}{dx} \\ &= \frac{1}{\Delta}\left(\sigma_{\mathrm{O}^{2-}} + 2\sigma_{\mathrm{O}^-}\right)\left(ej_\mathrm{charge}\right)\left[\left(\frac{\sigma_{\mathrm{O}^{2-}} + \sigma_\mathrm{e} + \sigma_\mathrm{h} + \sigma_{\mathrm{O}^-}}{\sigma_{\mathrm{O}^{2-}} + 2\sigma_{\mathrm{O}^-}}\right)\left(\frac{-2eJ_\mathrm{O}}{j_\mathrm{charge}}\right) - 1\right] \end{aligned} \tag{16}$$

At the steady state, $t_\mathrm{O} = \dfrac{-2eJ_\mathrm{O}}{j_\mathrm{charge}}$ is a constant, as are $J_\mathrm{O}$ and $j_\mathrm{charge}$. Here, the dimensionless $t_\mathrm{O}$ is always positive because in the SOEC mode, $J_\mathrm{O} > 0$ and $j_\mathrm{charge} < 0$, whereas in the SOFC mode, $J_\mathrm{O} < 0$ and $j_\mathrm{charge} > 0$.

The steady-state solution can now be obtained as follows. First, to solve $t_\mathrm{O}$ we integrate Eq. (16) from $x$=0 to $x$=$L$ to obtain

$$L = \int_{\mu'_{\mathrm{O}_2}}^{\mu''_{\mathrm{O}_2}} \frac{\left[\sigma_{\mathrm{O}^{2-}}\left(\sigma_\mathrm{e} + \sigma_\mathrm{h} + \sigma_{\mathrm{O}^-}\right) + 4\sigma_{\mathrm{O}^-}\left(\sigma_\mathrm{e} + \sigma_\mathrm{h}\right)\right] d\mu_{\mathrm{O}_2}}{4ej_\mathrm{charge}\left[\left(\sigma_{\mathrm{O}^{2-}} + \sigma_\mathrm{e} + \sigma_\mathrm{h} + \sigma_{\mathrm{O}^-}\right)t_\mathrm{O} - \left(\sigma_{\mathrm{O}^{2-}} + 2\sigma_{\mathrm{O}^-}\right)\right]} \tag{17}$$

This gives a unique $t_\mathrm{O}$ for each set of $j_\mathrm{charge}$ and boundary oxygen potentials, at $x$=0 and $L$. With this $t_\mathrm{O}$, the oxygen potential distribution can be obtained in terms of $j_\mathrm{charge}$ by integrating $\dfrac{d\mu_{\mathrm{O}_2}}{dx}$ from $x$=0 to an arbitrary $x$

$$x = \int_{\mu'_{O_2}}^{\mu_{O_2}} \frac{\left[\sigma_{O^{2-}}\left(\sigma_e+\sigma_h+\sigma_{O^-}\right)+4\sigma_{O^-}\left(\sigma_e+\sigma_h\right)\right]d\mu_{O_2}}{4ej_{charge}\left[\left(\sigma_{O^{2-}}+\sigma_e+\sigma_h+\sigma_{O^-}\right)t_O-\left(\sigma_{O^{2-}}+2\sigma_{O^-}\right)\right]} \quad (18)$$

Knowing the distribution of oxygen potential, we can next determine all the spatial distributions of conductivity. Then, Eq. (14) will provide $\frac{d\tilde{\mu}_{O^{2-}}}{dx}$ and $\frac{d\tilde{\mu}_e}{dx}$ at all $x$, and by integration, $\tilde{\mu}_{O^{2-}}$ and $\tilde{\mu}_e$. Finally, Eq. (7-10) will provide all the fluxes of individual species.

In the above, we considered the case of a steady-state $j_{charge}$ and $J_O$. However, under the open circuit voltage (OCV) condition when $j_{charge}$=0 and $t_O=\infty$, it is more convenient to instead solve the potential in terms of $J_O$, obtained from rewriting Eq. (16) and (17) in favor of $J_O$. This gives

$$L = \int_{\mu'_{O_2}}^{\mu''_{O_2}} \frac{\left[\sigma_{O^{2-}}\left(\sigma_e+\sigma_h+\sigma_{O^-}\right)+4\sigma_{O^-}\left(\sigma_e+\sigma_h\right)\right]d\mu_{O_2}}{-8e^2J_O\left(\sigma_{O^{2-}}+\sigma_e+\sigma_h+\sigma_{O^-}\right)} \quad (19)$$

and the oxygen potential distribution in terms of $J_O$ is given by

$$x = \int_{\mu'_{O_2}}^{\mu_{O_2}} \frac{\left[\sigma_{O^{2-}}\left(\sigma_e+\sigma_h+\sigma_{O^-}\right)+4\sigma_{O^-}\left(\sigma_e+\sigma_h\right)\right]d\mu_{O_2}}{-8e^2J_O\left(\sigma_{O^{2-}}+\sigma_e+\sigma_h+\sigma_{O^-}\right)} \quad (20)$$

The rest of the solution procedure is the same as before. The formulation of the problem is now complete.

## III. Oxygen Potential: No Defect association

In this section, we will first recover the Jacobsen-Mogensen solution for fully ionized lattice oxygen and oxygen defects without considering (ii) [25], then illustrate how the oxygen potential transition manifests under this condition for different current densities and electrode thickness. Without (ii), $\sigma_{O^-}=0$ and is dropped from Eq. (17-20). Thus, $t_O$ reduces to the transference number of ionic conduction, $t_i$. The Jacobsen-Mogensen solution is for a YSZ electrolyte of $L$=200 μm at 1000 °C between a hydrogen electrode (at $x$=0) and an oxygen electrode (at $x$=$L$), operated under either the SOFC mode ( $j_{\text{charge}} > 0$ ) or the SOEC mode ( $j_{\text{charge}} < 0$ ) with a modest current density or under the open circuit condition (OCV), and assuming $\sigma_{O^{2-}}$ independent of $\mu_{O_2}$ , and $\sigma_e \propto \exp\left(-\frac{\mu_{O_2}}{4RT}\right)$ and $\sigma_h \propto \exp\left(\frac{\mu_{O_2}}{4RT}\right)$ from standard defect chemistry consideration [24,25], which is consistent with **Fig. 1**. Using the conductivities in the figure, we obtained the oxygen potential distributions shown in **Fig. 3a**, which agree with **Fig. 5** and **7** of Ref. [25]. Jacobsen and Mogensen also considered the case of electrode overpotentials caused by limited interfacial reactions, which we too reproduced in our calculations (data not shown) by similarly accounting for them in the boundary conditions.

Regardless of mode and current density, solutions in **Fig. 3a** display a transition over a short distance (on the order of 10 μm), from a low oxygen potential on one side to a high oxygen potential on the other side. The inflection point of the transition is always around −2 eV, which corresponds to the potential where the minimum electronic conductivity lies at 1000 °C in **Fig. 1**. This is not coincidental. While the electronic current can be readily supported by the large electronic conductivity available at the high/low oxygen potential on the two sides, it is hampered by the minimum electronic conductivity in the middle, at about −2 eV, of **Fig. 1**. Since this potential lies between $\mu'_{O_2}$ = −3.79 eV (corresponding to $10^{-15}$ atm effective oxygen partial pressure on the hydrogen electrode; we set the oxygen potential at 1 atm oxygen partial pressure to be 0 eV) and $\mu''_{O_2}$ = −0.18 eV (corresponding to 0.2 atm effective oxygen partial pressure on the oxygen electrode) [25], it must be traversed and it holds the steepest oxygen potential gradient when forcing through a steady-state electronic current. Such is the origin of the oxygen potential transition.

Mathematically, the above can be easily verified by writing the explicit form of Eq. (17) and Eq. (19) with $\sigma_{\mathrm{O}^{-}} = 0$

$$L = \int_{\mu'_{\mathrm{O}_2}}^{\mu''_{\mathrm{O}_2}} \frac{d\mu_{\mathrm{O}_2}}{f(x)} \tag{21}$$

where the denominator $f(x)$

$$f(x) = \frac{d\mu_{\mathrm{O}_2}}{dx} = 2\frac{d\tilde{\mu}_{\mathrm{O}^{2-}}}{dx} - 4\frac{d\tilde{\mu}_{\mathrm{e}}}{dx} \tag{22}$$

is the slope of the oxygen potential. Specifically, we have

SOFC mode ( $j_{\mathrm{charge}} > 0$ , $t_{\mathrm{i}}>0$): $$f(x) = \frac{d\mu_{\mathrm{O}_2}}{dx} = 4ej_{\mathrm{charge}}\left(\frac{t_{\mathrm{i}}}{\sigma_{\mathrm{O}^{2-}}} + \frac{|t_{\mathrm{i}} - 1|}{\sigma_{\mathrm{e}} + \sigma_{\mathrm{h}}}\right) \tag{23a}$$

SOEC mode ( $j_{\mathrm{charge}} < 0$ , $t_{\mathrm{i}}<0$): $$f(x) = \frac{d\mu_{\mathrm{O}_2}}{dx} = 4ej_{\mathrm{charge}}\left(\frac{t_{\mathrm{i}}}{\sigma_{\mathrm{O}^{2-}}} - \frac{|1 - t_{\mathrm{i}}|}{\sigma_{\mathrm{e}} + \sigma_{\mathrm{h}}}\right) \tag{23b}$$

OCV ( $j_{\mathrm{charge}} = 0$ ): $$f(x) = \frac{d\mu_{\mathrm{O}_2}}{dx} = 4ej_0\left(\frac{1}{\sigma_{\mathrm{O}^{2-}}} + \frac{1}{\sigma_{\mathrm{e}} + \sigma_{\mathrm{h}}}\right) \tag{23c}$$

Here, in the expression for OCV, we used $j_0 = j_{\mathrm{O}^{2-}} = -(j_{\mathrm{e}} + j_{\mathrm{h}})$ to denote the absolute value of the ionic and electronic current densities. For example, take the case of SOEC mode where $j_{\mathrm{charge}}<0$, a monotonic positive slope is possible only if $t_{\mathrm{i}} \le 1$. (For SOFC, $j_{\mathrm{charge}}>0$, and $t_{\mathrm{i}} \ge 1$, the following argument also holds.) The slope is small on the two sides near $\mu'_{\mathrm{O}_2}$ and $\mu''_{\mathrm{O}_2}$ because of a large $\sigma_{\mathrm{e}} + \sigma_{\mathrm{h}}$, but steep near the minimum $\sigma_{\mathrm{e}} + \sigma_{\mathrm{h}}$. Clearly, the condition for a sharp potential transition is

$$L >> \frac{\mu''_{\mathrm{O}_2} - \mu'_{\mathrm{O}_2}}{[f(x)]_{\max}} \tag{24}$$

This criterion is satisfied by all the cases in **Fig. 3a**. For example, in the SOEC mode, the LHS of Eq. (24) =200 μm >> RHS=17 μm with −1 A/cm$^2$, and also >> RHS=5.7 μm with −3 A/cm$^2$.

Below, we will find this criterion also holds for other more general cases.

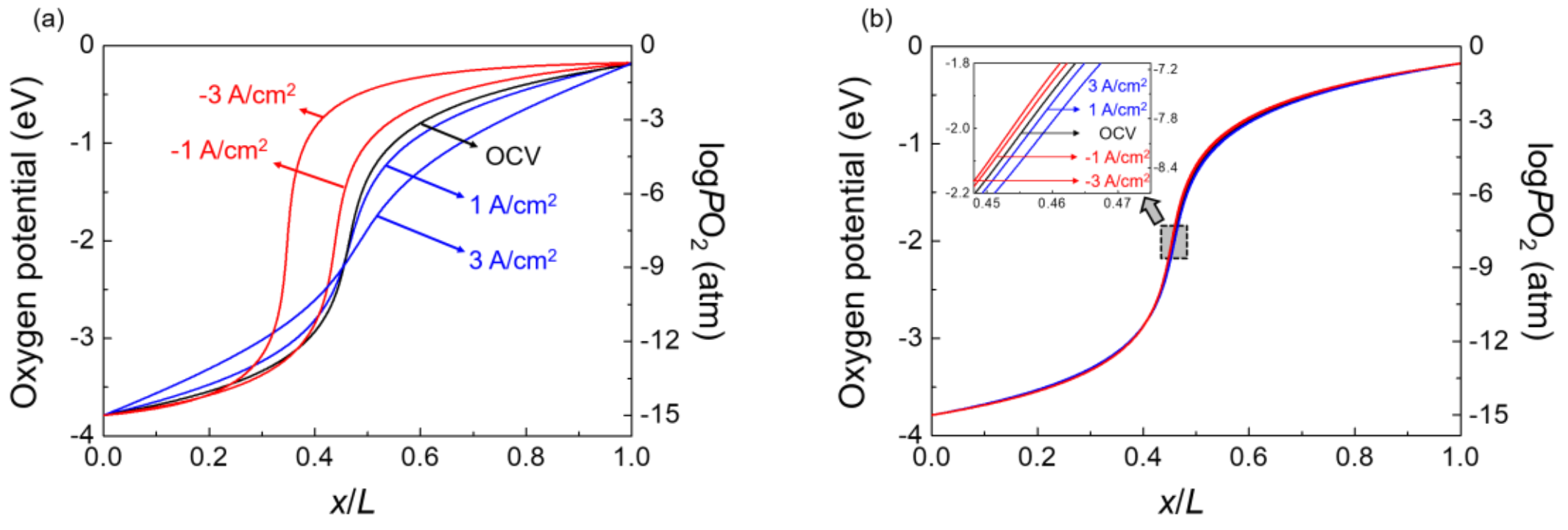

**Figure 3** (a) Calculated spatial distributions of oxygen potential without allowing defect association for SOFC modes (blue curves; total positive current densities indicated), SOEC modes (red curves; total negative current densities indicated) and OCV mode (black curve; zero total current density). (b) Same cases as (a) but now allowing internal reaction with $\alpha_e$=$\alpha_h$=0.475 (see Section V for details). Inset: enlarged view of center region. Temperature: 1000 °C, thickness: $L$=200 μm, oxygen potential from −3.79 eV to −0.18 eV along distance $x$ from left electrode.

The relatively thin $L$ used in the above calculations is appropriate for SOFC and SOEC. Obviously, Eq. (24) dictates that the transition should be even sharper in a thicker electrolyte under the same boundary potentials and current density, which is verified in **Fig. 4a**, for the SOEC mode. Remarkably, the figure also reveals a new feature not manifest in the Jacobsen-Mogensen solution [25]: As the thickness increases, the transition is increasingly shifted to the cathode. A similar trend is followed in **Fig. 4b** at a fixed thickness but with increasing current density. We shall refer this as cathode localization (of the transition). Anode localization is also possible but to avoid repetition we will postpone its discussion to a later section. The equivalence between increasing thickness and increasing current density can be understood by rewriting Eq. (7-10) as $j_i L = -\frac{\sigma_i}{n_i e}\frac{\partial \tilde{\mu}_i}{\partial (x/L)}$ ($n_i$ denoting the formal charge of species $i$), so the solutions with the same $j_i L$ and the same boundary oxygen potentials are the same when plotted against $x/L$.

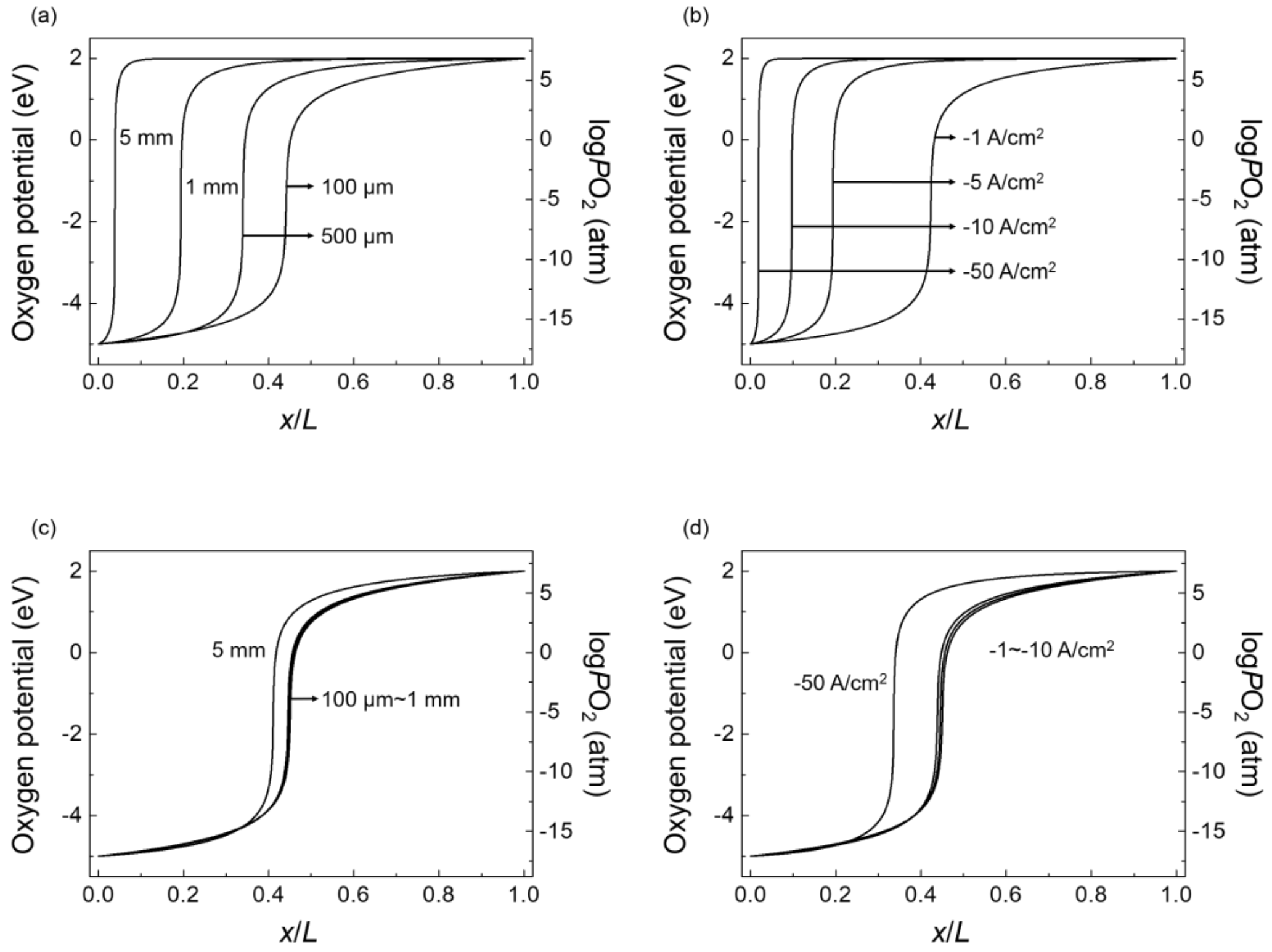

**Figure 4** Calculated spatial distributions of oxygen potential in SOEC mode without allowing defect association reactions under (a) fixed current density of −5 A/cm$^2$ and varying electrolyte thickness from 100 μm to 5 mm, and (b) fixed electrolyte thickness of 1 mm and varying current density of −1 A/cm$^2$ to −50 A/cm$^2$. (c) Same cases as (a), and (d) same cases as (b), but now allowing internal reaction with $\alpha_e$=$\alpha_h$=0.475 (see Section V for details). Temperature: 1250 $^o$C, oxygen potential from −5 eV to 2 eV along distance $x$ from left electrode. Note: Hole conductivity at 2 eV (0.032 $\Omega^{-1}$cm$^{-1}$) is larger than electron conductivity at −5 eV (0.026 $\Omega^{-1}$cm$^{-1}$) according to **Fig. 1**, which causes transition to localize at cathode.

We conclude this section by comparing the problem of oxygen potential distribution in mixed conducting electrolyte and the problem of temperature distribution in solidification of a pure substance, both in one dimension. Because the sharp transition is always fixed at the characteristic oxygen potential $\mu^*_{O_2}$ at the minimum $\sigma_e + \sigma_h$, we may recognize the following analogy where "≈" means "corresponds to": The high potential side ≈ the liquid-phase high-temperature side, the low potential side ≈ the solid-phase low-temperature side, the transition potential $\mu^*_{O_2}$ ≈ the melting point $T_m$, and the boundary potentials $\mu'_{O_2} < \mu^*_{O_2} < \mu''_{O_2}$ ≈ the temperatures $T' < T_m < T''$. In steady-state solidification, the heat flux is constant throughout $L$, which uniquely determines the location of the solid/liquid interface. For example, if $\kappa_{liquid}\left(T'' - T_m\right) >> \kappa_{solid}\left(T_m - T'\right)$, where $\kappa$ is the thermal conductivity, then the solid-liquid interface is very near the left end, which is analogous to the case in **Fig. 4a-b**. ($\sigma_h$ at $\mu''_{O_2}$ is larger than $\sigma_e$ at $\mu'_{O_2}$.) Thus the analogy is very useful for understanding the essential physics of our problem. However, our problem having two independent potentials is more complicated than the solidification problem of a pure substance with only one field variable, the temperature. Therefore, we need to specify not only the boundary potentials but also the flux, which can be $j_{charge}$, $J_O$, or their linear combination. This is the essence of the problem. The non-linear conductivity in **Fig. 1** makes finding the solution more cumbersome and the transition sharper but does not fundamentally alter the nature of the problem.

## IV. Solution to the Park-Blumenthal Problem

To illustrate how $O^-$ diffusion may manifest in real applications, we analyze below the celebrated Park-Blumenthal experiment [24] using Eq. (17-20). The experiment measured electronic conductivities using a sealed chamber, placed in an environment of $p_{O_2}^{out}$ and bounded by two electrically isolated YSZ membranes, each of a thickness $L$. To establish an oxygen potential gradient, they extracted oxygen by passing a current density $j_{charge}$ through one membrane in the SOEC mode. To monitor the oxygen pressure $p_{O_2}^{in}$ inside, they kept the other membrane in the OCV mode to define the Nernst voltage, $V = \frac{k_B T}{4e} \ln \frac{p_{O_2}^{out}}{p_{O_2}^{in}} = \frac{\mu''_{O_2} - \mu'_{O_2}}{4e}$. This provides two measurables, $j_{charge}$ and $\mu''_{O_2} - \mu'_{O_2}$, which are solved in the limit of $\mu''_{O_2} \approx \mu'_{O_2}$. In this limit, the integrand of Eq. (17) for the SOEC membrane equals the integrand of Eq. (19) of the OCV membrane, giving

$$4ej_{charge}\left[\left(\sigma_{O^{2-}} + \sigma_e + \sigma_h + \sigma_{O^-}\right)t_O - \left(\sigma_{O^{2-}} + 2\sigma_{O^-}\right)\right] = -8e^2 J_O\left(\sigma_{O^{2-}} + \sigma_e + \sigma_h + \sigma_{O^-}\right) \quad (25)$$

At the steady state, $J_O$ is the same in both membranes. So, from Eq. (25) and $J_O = \frac{t_O}{-2e} j_{charge}$, we have

$$t_O = \frac{\sigma_{O^{2-}} + 2\sigma_{O^-}}{2\left(\sigma_{O^{2-}} + \sigma_e + \sigma_h + \sigma_{O^-}\right)} \quad (26)$$

Note $t_O \approx 0.5$ in the limit of $\sigma_{O^{2-}} >> \max\left\{\sigma_e, \sigma_h, \sigma_{O^-}\right\}$, which is the same conclusion from the equivalent circuit analysis without $O^-$ conduction as illustrated in the **Appendix**. ($t_O = 0.5$ implies the charge current is equally shared by the ionic current and the electronic current.) Substituting $t_O$ into Eq. (17) in the same limit, we obtain

$$\begin{aligned} -\frac{1}{2} j_{charge} &= \left\{\frac{\sigma_{O^{2-}}\left(\sigma_e + \sigma_h + \sigma_{O^-}\right) + 4\sigma_{O^-}\left(\sigma_e + \sigma_h\right)}{\sigma_{O^{2-}} + 2\sigma_{O^-}}\right\} \frac{\mu''_{O_2} - \mu'_{O_2}}{4e} \frac{1}{L} \\ &\approx \left(\sigma_e + \sigma_h + \sigma_{O^-}\right) \frac{\mu''_{O_2} - \mu'_{O_2}}{4e} \frac{1}{L} \end{aligned} \quad (27)$$

In the second equality above, the approximation of $\sigma_{O^{2-}} >> \max\left\{\sigma_e, \sigma_h, \sigma_{O^-}\right\}$ was made. Therefore, contrary to the claim of Park and Blumenthal, their experiment actually measured $\sigma_e + \sigma_h + \sigma_{O^-}$ instead of $\sigma_e + \sigma_h$.

The reason that $\sigma_{O^-}$ cannot be separated from $\sigma_e$ and $\sigma_h$ in the experiment can be seen in the schematic of **Fig. 5**, which illustrates three closed circuit loops in the OCV membrane to transport $O_2$ into the chamber. Since these loops all contribute to the current, their conductivities are additive and they all belong to the "electronic conductivity" measured by Park and Blumenthal. That is, the sum of the red and blue curves in **Fig. 1** is not $\sigma_e + \sigma_h$ but rather $\sigma_e + \sigma_h + \sigma_{O^-}$. This is why in the caption of **Fig. 1**, we call the left branch (red) $\sigma_e$* so it may include both $\sigma_e$ and $\sigma_{O^-}$, and the right branch (blue) $\sigma_h$* so it may include both $\sigma_h$ and $\sigma_{O^-}$.

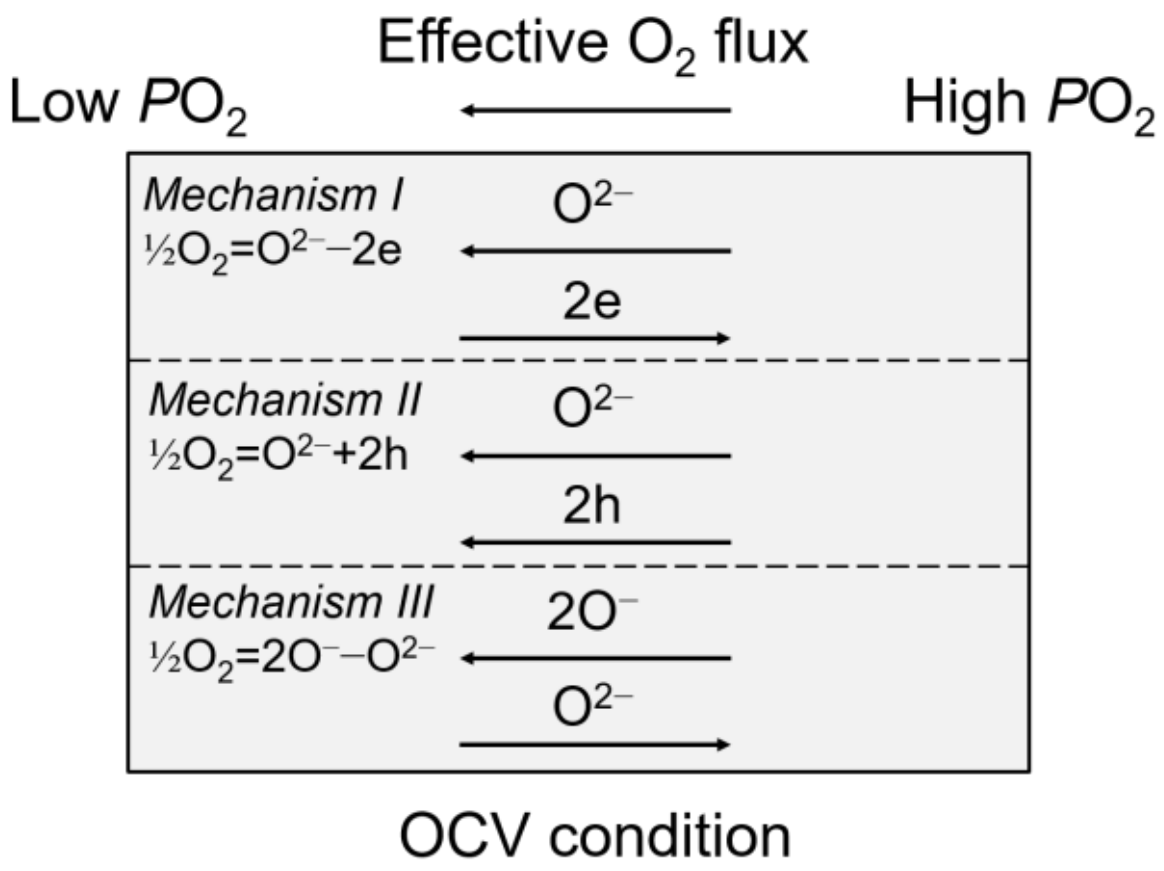

**Figure 5** Schematic $O^{2-}$, $O^-$, e and h transport under OCV condition, providing three parallel mechanisms for $O_2$ transport. In each, overall kinetics is limited by the slower diffusing species. Together, it is controlled by the fastest of the three slower species.

## V. Oxygen Potential: with Internal Reactions

We have associated $O^-$ diffusion to either an electron-mediated mechanism—a lattice $O^{2-}$ exchanges position with an electron-tagged oxygen vacancy $\left(V_O^{\bullet\bullet} + e\right)$, or a hole-mediated mechanism—an oxygen vacancy exchanges position with a hole-tagged lattice oxygen $\left(O^{2-} + h\right)$, and both mechanisms may contribute to the electronic conductivity $\sigma_e + \sigma_h + \sigma_{O^-}$ in

**Fig. 1**. Following the notation of $\sigma_e$* and $\sigma_h$* above and setting the total electronic conductivity the same as Park and Blumenthal's measurement, $\sigma_e + \sigma_h + \sigma_{O^-} = \sigma_e^* + \sigma_h^*$, we write

$$\sigma_e = (1 - \alpha_e)\sigma_e^* \qquad (28a)$$

$$\sigma_h = (1 - \alpha_h)\sigma_h^* \qquad (28b)$$

$$\sigma_{O^-} = \alpha_e \sigma_e^* + \alpha_h \sigma_h^* \qquad (28c)$$

Below, for simplicity, $\alpha_e$ and $\alpha_h$ are treated as model parameters independent of oxygen potential, and the mobility of $V_O^{\bullet\bullet}$ and $O^{2-}$ was not considered because in YSZ, they are not rate-limiting having a substantially lower activation energy than those of electrons and holes.

Adopting Eq. (28), we perform the following calculations summarized in **Fig. 6**. Starting with $\alpha_e = \alpha_h = 0$, which corresponds to not having any $O^-$, we recover in **Fig. 6a-b** (shown as the black curve) the same results as in **Fig. 4a-b** of cathode localization (of potential transition). Cathode localization remains intact when only the electron-mediated mechanism is switched on ($\alpha_e > \alpha_h = 0$, **Fig. 6a**). Conversely, anode localization takes over when only the hole-mediated mechanism is switched on ($\alpha_h > \alpha_e = 0$, **Fig. 6b**). Further intermediate cases featuring a sharp crossover from cathode localization to anode localization manifest in **Fig. 6c-d** when $O^-$ diffusion crossovers from electron dominating ($\alpha_e > \alpha_h > 0$) to hole dominating ($\alpha_h > \alpha_e > 0$). Therefore, there is a definite association between cathode localization and electron-mediated mechanism, vs. anode localization and hole-mediated mechanism.

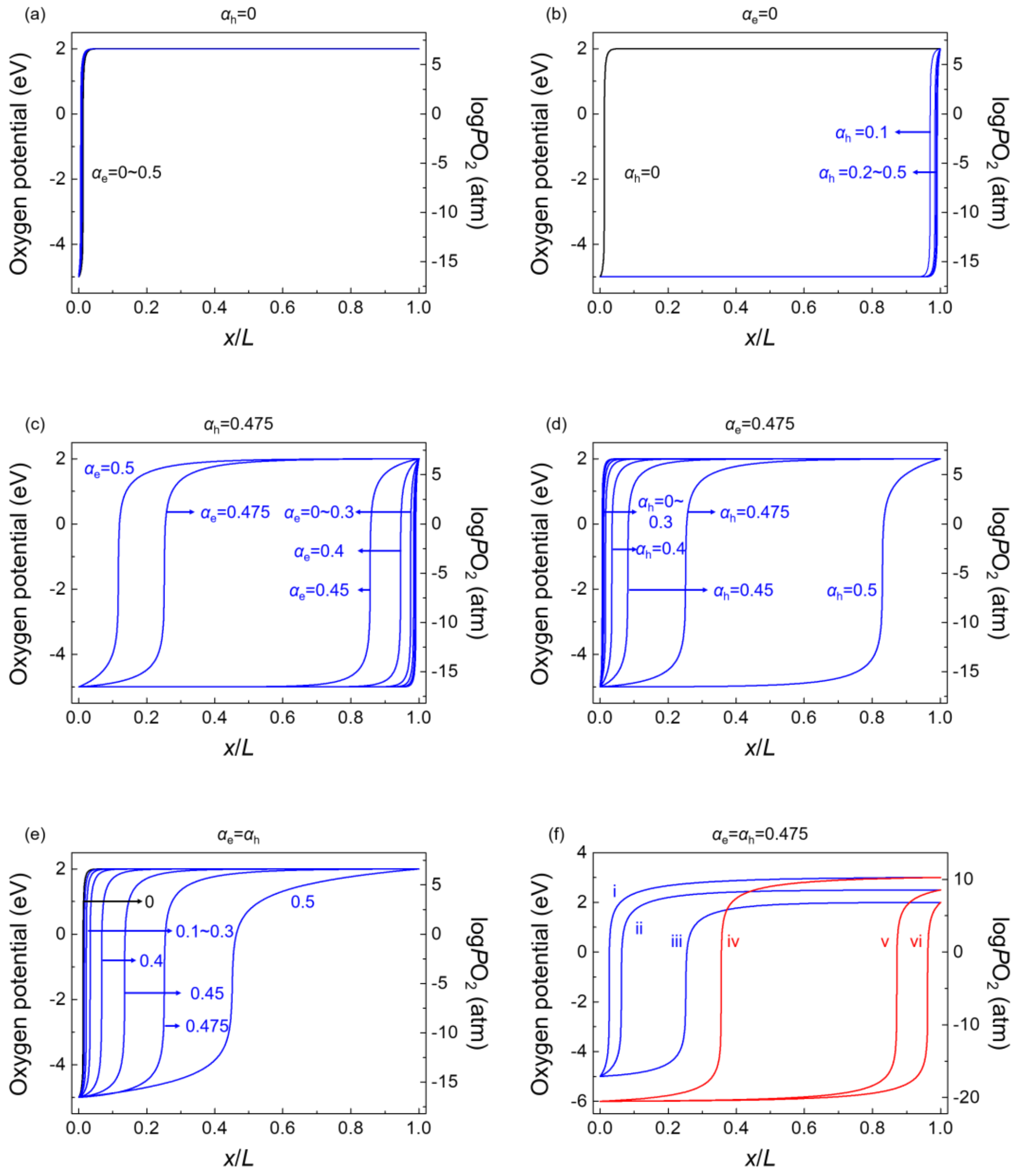

**Figure 6** Calculated spatial distributions of oxygen potential in SOEC mode with (a) $\alpha_h$=0, $\alpha_e$ from 0 to 0.5; (b) $\alpha_e$=0, $\alpha_h$ from 0 to 0.5; (c) $\alpha_h$=0.475, $\alpha_e$ from 0 to 0.5; (d) $\alpha_e$=0.475, $\alpha_e$ from 0 to 0.5; (e) $\alpha_e$=$\alpha_h$ ranging from 0 to 0.5 as marked. Oxygen potential from −5 eV to 2 eV along distance $x$ from left electrode. (f) Calculated spatial distributions of oxygen potential with $\alpha_e$=$\alpha_h$=0.475, under different oxygen potential ranges; i: from −5 eV to 3 eV, ii: from −5 eV to 2.5 eV, iii: from −5 eV to 2 eV, iv: from −6 eV to 3 eV, v: from −6 eV to 2.5 eV and vi: from −6 eV to 2 eV, all along distance $x$ from left electrode. Temperature: 1250 °C, thickness: $L$=1.5 mm,

current density: −50 A/cm$^2$.

An interesting new phenomenon emerges when $\alpha_e \approx \alpha_h > 0$. This is illustrated in **Fig. 6c-d** where the potential transition falls far away from either electrode. To better illustrate the trend, we present in **Fig. 6e** the cases for $\alpha_e = \alpha_h$ increasing from 0 to 0.5. Here, $O^-$ diffusion is aided equally by electron-mediated and hole-mediated mechanisms (despite $[e] \neq [h]$ in general), and as the role of $O^-$ diffusion becomes more prominent, the transition gradually shifts to the center. Another interesting effect of $O^-$ diffusion is illustrated in **Fig. 3b**, where we repeat the calculations for the problem stated for **Fig. 3a** but now allowing $O^-$ diffusion at $\alpha_e = \alpha_h = 0.475$. Remarkably, the oxygen potential distribution with a transition near the center becomes almost completely independent of the mode of operation and the current density, unlike the case in **Fig. 3a**. The same effect is further illustrated in **Fig. 4c-d,** which repeats the calculations of oxygen potential for the problem stated for **Fig. 4a-b**—again allowing $O^-$ diffusion: Unlike in **Fig. 4a-b**, the distributions in **Fig. 4c-d** with a transition near the center only very weakly depend on *L* and current density. Therefore, the profound effects of electron-mediated and hole-mediated $O^-$ conductivity on potential transition are verifiable in more than one way.

## VI. Oxygen Potential Transition: A Closer Look

The transitions in **Fig. 4c-d** and **Fig. 6a-e** again occur at the potential at the minimum of the combined red-blue curve in **Fig. 1**, which as we now know corresponds to the minimum of $\sigma_e + \sigma_h + \sigma_{O^-}$. This is because when $\sigma_{O^{2-}} >> \sigma_e + \sigma_h + \sigma_{O^-}$, Eq. (16) reduces to the following to a first approximation

$$\frac{d\mu_{O_2}}{dx} = \frac{4\left(-2e^2 J_O - ej_{charge}\right)}{\sigma_e + \sigma_h + \sigma_{O^-}} \qquad (29a)$$

Therefore, at the steady state when both $J_O$ and $j_{charge}$ are constant across $L$, the steepest slope of oxygen potential must coincide with the minimum of $\sigma_e + \sigma_h + \sigma_{O^-} = \sigma_e^* + \sigma_h^*$ in **Fig. 1**. This is the same condition established in **Section III**, so the origin of the potential transition remains the same with and without internal reactions. It also follows that if $\frac{d\mu_{O_2}}{dx}$ is very large at the transition, then for larger $L$ that satisfies Eq. (24), $\frac{d\mu_{O_2}}{dx}$ elsewhere must be very small. This is possible if and only if the oxygen potential approaches the two boundary values, $\mu'_{O_2}$ and $\mu''_{O_2}$. That is, there is a step-like transition from one boundary value to the other.

The same argument also explains why the transition is localized near one electrode. To keep $\frac{d\mu_{O_2}}{dx}$ very small over a very long distance across $L$, most of the length must be spent near the oxygen potential where $\sigma_e^* + \sigma_h^*$ is the largest. In the range of −5 eV to 2 eV at 1,250 °C shown in **Fig. 4**, this falls at 2 eV, which explains why there is a wider flat region at such potential in the figure, hence cathode localization therein. Obviously, it is entirely possible to "switch side" to anode localization by adjusting boundary potentials so that the anode side offers a larger $\sigma_e^* + \sigma_h^*$. This is illustrated in **Fig. 6f**. The switch-over can be quite abrupt, e.g., from curves (ii) to (v) merely requires lowering the cathodic oxygen potential by 1 eV; likewise, switching from (iv) to (v) is triggered by lowering the anodic oxygen potential by 0.5 eV.

Abrupt side-switching also happens upon a small change in $\alpha_{e/h}$. Again, this can be explained by Eq. (16), but higher order terms of the order of $\left(\sigma_e + \sigma_h + \sigma_{O^-}\right)/\sigma_{O^{2-}}$ now need to be included to extend Eq. (29a). Noting in Eq. (15), $\Delta \approx \sigma_{O^{2-}}\left(\sigma_e + \sigma_h + \sigma_{O^-}\right)$ within the same order of approximation, we rewrite Eq. (16) in two equivalent forms

$$\frac{d\mu_{O_2}}{dx}=\frac{4\left(\sigma_{O^{2-}}+2\sigma_{O^-}\right)\left(ej_{charge}\right)}{\sigma_{O^{2-}}\left(\sigma_e+\sigma_h+\sigma_{O^-}\right)}\left[\left(1+\frac{\sigma_e+\sigma_h-\sigma_{O^-}}{\sigma_{O^{2-}}}\right)\left(\frac{-2eJ_O}{j_{charge}}\right)-1\right] \quad (29b)$$

$$\frac{d\mu_{O_2}}{dx}=\frac{4\left(\sigma_{O^{2-}}+2\sigma_{O^-}\right)\left(ej_{charge}\right)}{\sigma_{O^{2-}}\left(\sigma_e+\sigma_h+\sigma_{O^-}\right)}\left[\frac{\sigma_e+\sigma_h-\sigma_{O^-}}{\sigma_{O^{2-}}}-\frac{j_e+j_h-j_{O^-}}{j_{charge}}\right] \quad (29c)$$

This result (especially Eq. (29b)) allows us to find the side where the flattest potential lies, i.e., where $\frac{d\mu_{O_2}}{dx}$ is minimally positive. There are two pertinent considerations here. (a) $\left|\frac{d\mu_{O_2}}{dx}\right|$ in Eq. (29b-c) is minimized at $\left(\sigma_e+\sigma_h+\sigma_{O^-}\right)_{max}$, whose location is insensitive to the value of $\alpha_{e/h}$ and is mostly determined by $\mu'_{O_2}$ and $\mu''_{O_2}$. (Here, "$()_{max}$" refers to the global maximum of the bracketed quantity in the entire thickness.) (b) The square bracket on the right-hand side of Eq. (29b) is in some cases (see below) minimized at $\left(\sigma_e+\sigma_h-\sigma_{O^-}\right)_{max}$, whose location can abruptly pivot between $\mu'_{O_2}$ and $\mu''_{O_2}$ upon small changes in $\alpha_{e/h}$. For example, in the SOEC mode where $J_O>0$ and $j_{charge}<0$, the square bracket on the right-hand side of Eq. (29b) is negative and $\frac{-2eJ_O}{j_{charge}}$ is positive, so maximizing $\sigma_e+\sigma_h-\sigma_{O^-}$ will minimize $\frac{d\mu_{O_2}}{dx}$. In fact, in this mode, we find (b) is given by $\left(\sigma_e+\sigma_h-\sigma_{O^-}\right)_{max}=\left[\left(1-2\alpha_e\right)\sigma_e^*+\left(1-2\alpha_h\right)\sigma_h^*\right]_{max}$, which can be shown to correctly explain all the "side-switching" phenomena in **Fig. 6**. Lastly, our earlier observation of an association between cathode localization in **Fig. 6** and electron-mediated mechanism is also understood: In electron-mediated $O^-$ diffusion, $\sigma_{O^-}\sim\sigma_e$, so $\left(\sigma_e+\sigma_h-\sigma_{O^-}\right)\sim\sigma_h$; therefore, $\left(\sigma_e+\sigma_h-\sigma_{O^-}\right)_{max}$ is located at the high potential end, hence cathode localization. Likewise, anode localization associated with hole-mediated mechanism can be explained.

Thus far, we have not checked whether $J_O$ is influenced by having $O^-$ diffusion.

Intuitively, we expect it to increase if some oxygen can latch onto electrons or holes to gain mobility. An increase in $\left|J_{\rm O}\right|$ (i.e., an increase in $t_{\rm O}$) is indeed verified in **Fig. 7a**, and for completeness, potentials of $O_2$, e, $O^{2-}$ and $O^{-}$, and conductivities of e+h, $O^{2-}$ and $O^{-}$ are also shown in **Fig. 7b-c**. Here, it is interesting to note that, at the transition, $j_{\rm O^{2-}}$ peaks to compensate for the $j_{\rm O^{-}}$ minimum, which is suppressed by the very low electronic conductivity. Since this peak $j_{\rm O^{2-}}$ is at a level much higher than the base line (broken black line, which is $j_{\rm O^{2-}}$ in the absence of internal reactions) whereas $j_{\rm O^{-}}$ is nearly zero (actually slightly negative at the transition as seen from **Fig. 7b**), enhanced $\left|J_{\rm O}\right|$ is verified at the transition to which $j_{\rm O^{2-}}$ is the main contributor. However, to verify the overall $\left|J_{\rm O}\right|$ enhancement, one need to inspect the two (electrode) ends, where the sum of $j_{\rm O^{2-}}$ and $j_{\rm O^{-}}$ again much exceeds the base line, and it is $j_{\rm O^{-}}$, helped by the largest $\sigma_{\rm e}$ or $\sigma_{\rm h}$ there, that is the main contributor to the enhanced $\left|J_{\rm O}\right|$. This makes it clear that the origin of enhanced $\left|J_{\rm O}\right|$ is electronic conduction—which is at maximum at the two ends—and defect association that enables ion-electron/hole coupling there. The enhancement of $\left|J_{\rm O}\right|$ is also seen in other cases. As $L$ increases or $\left|j_{\rm charge}\right|$ increases, and an increasing section of the length is spent at very near the boundary potentials to maximize electronic conductivity, it also benefits and maximizes $j_{\rm O^{-}}$. This allows the $\left|J_{\rm O}\right|$ enhancement to maintain and the peak $j_{\rm O^{2-}}$ at the transition to persist. It has a buffering effect on the transition: Because $j_{\rm charge}$ can also be carried by enhanced $j_{\rm O^{2-}}$, there is less need for $j_{\rm e}+j_{\rm h}$ (see **Fig. 7a**), hence less need to have a sharp transition in the potentials, including the tendency of having an increasingly larger $\frac{d\mu_{\rm O_2}}{dx}$ at the transition in **Fig. 4**. It may also be understood by looking again at the SOEC calculations in **Fig. 6.** Recall that in this mode, $J_{\rm O}>0$, $j_{\rm charge}<0$, $-2e^2J_{\rm O}-ej_{\rm charge}$ on the right-hand side of Eq. (29a) is positive, and the square bracket on the right-hand side of Eq. (29b) is negative with $\frac{-2eJ_{\mathrm{O}}}{j_{\mathrm{charge}}}$ being positive. So, an increase in $|J_{\mathrm{O}}|$ will cause (positive) $\frac{d\mu_{\mathrm{O_2}}}{dx}$ in Eq. (29a-b) to decrease at the transition, which is evident in **Fig. 6**. As a result, there is a lessened need for $\frac{d\mu_{\mathrm{O_2}}}{dx}$ elsewhere to maintain as flat as it was, which lessens the need to localize. All the above results are the consequence of enhanced oxygen-ion transport that results from ion-electron/hole reactions.

Lastly, although the above sample calculations were performed for a certain thickness and a fixed current density, all the results can be applied to other thickness and current density combinations when the potential distribution is plotted in $x/L$ and when $j_iL$ is the same. This is because as previously noted Eq. (7-10) can be written in the form of $j_i L = -\frac{\sigma_i}{n_i e}\frac{\partial \tilde{\mu}_i}{\partial (x/L)}$.

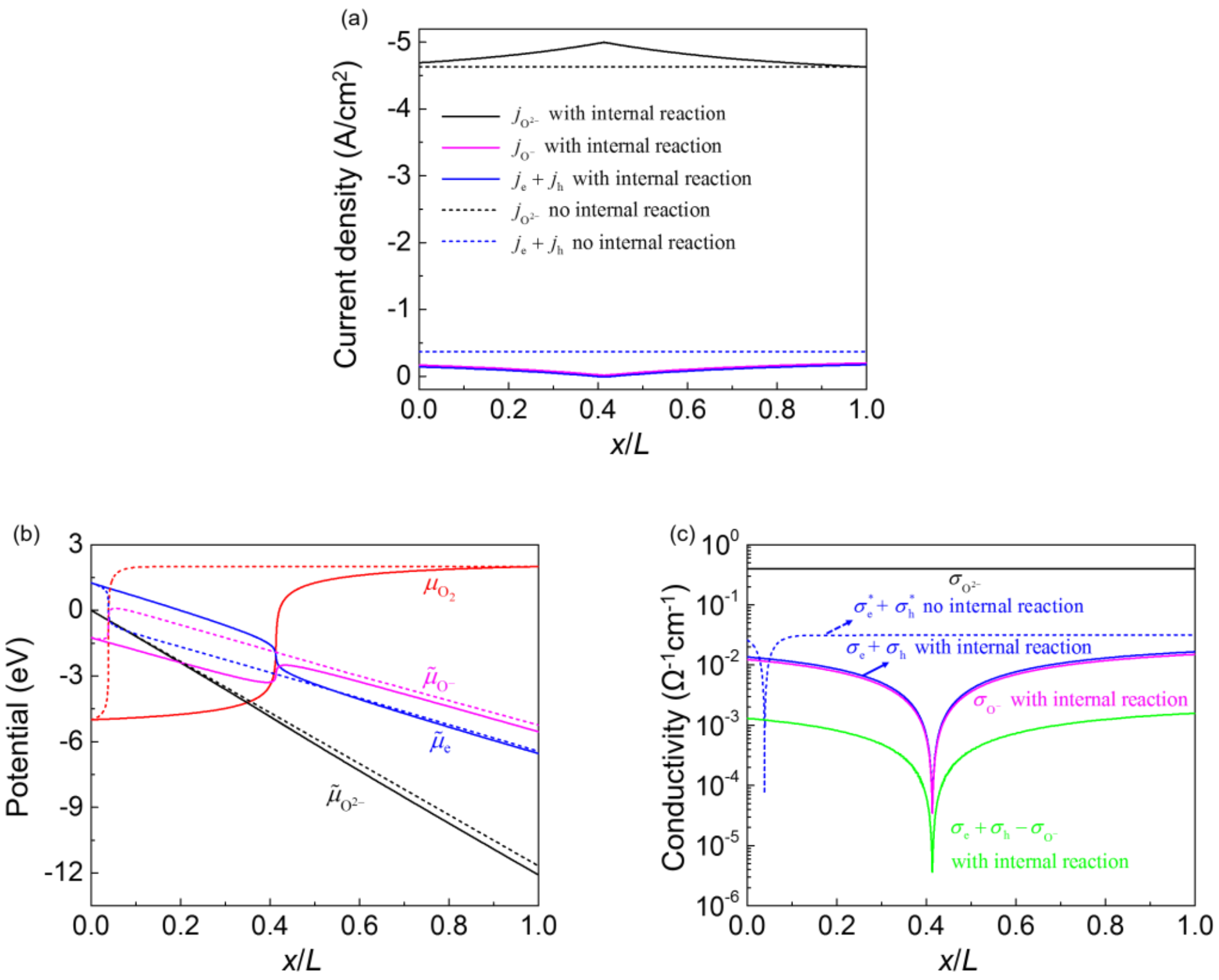

**Figure 7** Calculated (a) $O^{2-}$, $O^{-}$ and electronic current densities, (b) potential profiles for $O_2$, e,

$O^{2-}$ and $O^-$, and (c) conductivities of e+h, $O^{2-}$ and $O^-$ across the electrolyte in two cases: no internal reaction ($\alpha_e$=$\alpha_h$=0) shown by dashed curves, and with internal reaction with $\alpha_e$=$\alpha_h$=0.475 shown by solid curves. Temperature: 1250 °C, thickness: $L$=5 mm, oxygen potential from −5 eV to 2 eV along distance $x$ from left electrode, total current density: −5 A/cm$^2$. Same SOEC conditions as 5-mm case in **Fig. 4a** and **c**.

## VII. Discussion

(1) Why do we need to consider $O^-$ diffusion?

We have proposed to include $O^-$ diffusion in polarization modeling based on the idea that oxygen ions and oxygen vacancies can change their charge states by associating electrons and holes. This new concept is needed to reconcile a major discrepancy between our experimental observation on grain size transition and the previous calculations that did not consider $O^-$ diffusion. As illustrated by the example in **Fig. 2** and more in Ref. [14-17], at a sample temperature of 1200-1400 °C, a current density from −10 to −50 A/cm$^2$, and a sample thickness of ~1 mm, there is a very sharp grain size transition halfway across the electrically loaded YSZ, indicating a correspondingly sharp oxygen potential transition at the same position. (We proved the correlation between fast grain growth kinetics and depressed oxygen potential in Ref. [15].) This was not an isolated observation, because we had tested many samples of various thickness (but always more than 0.5 mm) with various terminal oxygen potentials $\mu'_{O_2}$ and $\mu''_{O_2}$, and always found the grain-size transition located at $x$~$L$/2. In contrast, not including $O^-$ diffusion, the Jacobsen-Mogensen solution [25] when performed for a 1 mm electrolyte always yields a transition localized at one of the electrodes (at $x$~0 or $L$) rather than at $x$~$L$/2. (See the black curve in **Fig. 6a** and **b**, with $\alpha_e$=$\alpha_h$=0.) Therefore, although it suffices in SOFC and SOEC devices that have thin electrolytes, the Jacobson-Mogenson solution lacks the full physics to be applicable in thicker electrolytes tested in our experiments. This discrepancy was resolved in the present study by considering $O^-$ diffusion.

To see a grain size transition in YSZ, a large current and high temperature is needed. Our grain size studies were typically conducted at above 1200 °C since there is very little grain boundary mobility in YSZ below this temperature [37]. Moreover, since YSZ is relatively difficult to reduce, we also used very large current densities to force huge electrode overpotentials to trigger grain growth. But a smaller grain size and another electrolyte (e.g., Gd-doped ceria, which is easier to reduce with a grain growth kinetics very sensitive to oxygen potential [15]) should allow the same observations at a lower temperature and a smaller current density. Indeed, similar grain size transition was also seen in flash-sintered ZnO (**Fig. 5** of Ref. [38]) and may exist in other flash-sintered ceramics. Other observations consistent with our calculations to be discussed next include the activation energies of electron and hole conductivity (see Subsection (2)), the lack of transition when the boundary potentials are all very negative (see Subsection (2)) and instability of grain size transition (see Subsection (3)).

(2) Electron/hole association

To describe $O^-$ diffusion, we have used the association parameters $\alpha_e$ and $\alpha_h$ for $\left(V_O^{\bullet\bullet}+e\right)$ and $\left(O^{2-}+h\right)$ complexes, respectively; without $O^-$, $\alpha_e$=$\alpha_h$=0. (As it will become clear later in this subsection, this is but one form of internal reactions between electronic species and ionic species.) Evidence for internal reactions between lattice defects and electrons/holes was already reviewed in **Introduction**. From **Fig. 6**, it is clear that a substantial $\alpha_e$ or $\alpha_h$ is needed for an appreciable effect of $O^-$ diffusion, hence to explain our experimental results summarized in the previous subsection. This in turn requires a large degree of association of oxygen vacancy with electron (e.g., $\left(V_O^{\bullet\bullet} + e\right)$ complex), or lattice oxygen with hole (e.g., $\left(O^{2-} + h\right)$ complex). The strongest evidence for such association is from the activation energies of $\sigma_e^*$, $\sigma_h^*$, and $\sigma_{O^{2-}}$, which are all comparable. To further support this, we have also calculated their mobilities, and found them of the same order of magnitude. (At 1000 ºC, $M_{O^{2-}} = 1.68 \times 10^{-4}$ $cm^2V^{-1}s^{-1}$, $M_e = 2.64 \times 10^{-5}$ $cm^2V^{-1}s^{-1}$ and $M_h = 5.92 \times 10^{-5}$ $cm^2V^{-1}s^{-1}$ ; at 1250 ºC, $M_{O^{2-}} = 5.49 \times 10^{-4}$ $cm^2V^{-1}s^{-1}$ , $M_e = 4.46 \times 10^{-4}$ $cm^2V^{-1}s^{-1}$ and $M_e = 2.85 \times 10^{-4}$ $cm^2V^{-1}s^{-1}$ .) Note that if there were no association/trapping at lattice/defect sites, this result would be very surprising because the mobilities of electron and hole should have been much larger than that of oxygen ion. But a strong association is consistent with the glassy energy landscape (analogous to a "compositional glass") of YSZ, which offers many lattice sites for possible electron/hole association. [39,40] It is also consistent with the observation that the activation energy of oxygen diffusion varies from 0.5 eV above 1000 ºC, to 0.79 eV at lower temperature, because a stronger association is expected at lower temperature when the configurational entropy against association is less important. These observations support our opinion that the data in **Fig. 1**, which has been attributed to electrons and holes in the past, is likely to have a substantial contribution from $O^-$ or the like, especially at lower temperature; i.e., $\alpha_e$ and $\alpha_h$ increases at lower temperature. However, the fact that the two branches in **Fig. 1** maintain their respective $PO_2^{1/4}$ and $PO_2^{-1/4}$ dependence as expected from defect chemistry consideration suggests that $\alpha_e$ and $\alpha_h$ are not strongly $PO_2$ dependent. This is reasonable since the concentrations of electrons and holes are usually rather low compared to that of oxygen vacancies.

Obviously, the highest electron and hole concentrations are near the two ends where the association of electron/hole with $O^{2-}/V_O^{\bullet\bullet}$ is likely to be most significant. Conversely, at the center section where the electron and hole concentrations are very small, association is unlikely to be important. Phenomenologically, this may be modeled by letting $\alpha_e$ and $\alpha_h$ be a function of oxygen potential; specifically, $\alpha_e$ becomes large at highly negative $\mu_{O_2}$ while $\alpha_h$ becomes large at highly positive $\mu_{O_2}$, but both are negligible elsewhere. This is illustrated in the inset of **Fig. 8** by using

$$\alpha_e = \frac{0.45}{1+\exp\left(\frac{\mu_{O_2}+4}{0.25}\right)} \quad (30a)$$

$$\alpha_h = \frac{0.45}{1+\exp\left(-\frac{\mu_{O_2}-1}{0.25}\right)} \quad (30b)$$

to represent such variation. With such variation, the oxygen potential distributions in Fig. 8 for the two reference cases (**I**: $\alpha_e$=$\alpha_h$=0, solid curve in black; and **II**: $\alpha_e$=$\alpha_h$=0.45, solid curve in purple) are changed into that for **Case III** (with variable $\alpha_e$ and $\alpha_h$, dashed curve in purple). Remarkably, even though the electron/hole association with $O^{2-}/V_O^{\bullet\bullet}$ in the above model is very limited ($\alpha_e$ and $\alpha_h$ << 0.45) except near the two ends, it gives **Case III** and can more effectively remove localization from the electrode ends than even **Case II** where more extensive defect association reactions ($\alpha_e$=$\alpha_h$=0.45) are allowed throughout the thickness. Therefore, the model has provided further support to our finding in **Section VI**: While the transition is always located at the conductivity minimum, localization can be very effectively alleviated by including the contribution of $j_{O^-}$ to $|J_O|$, which is most significant near the two electrodes where it is helped by the largest $\sigma_e$ or $\sigma_h$ there. In short, mere defect association near the electrodes has a profound effect on polarization distribution, and given the common occurrence of inefficient electrode kinetics we believe such reaction is all but inevitable.

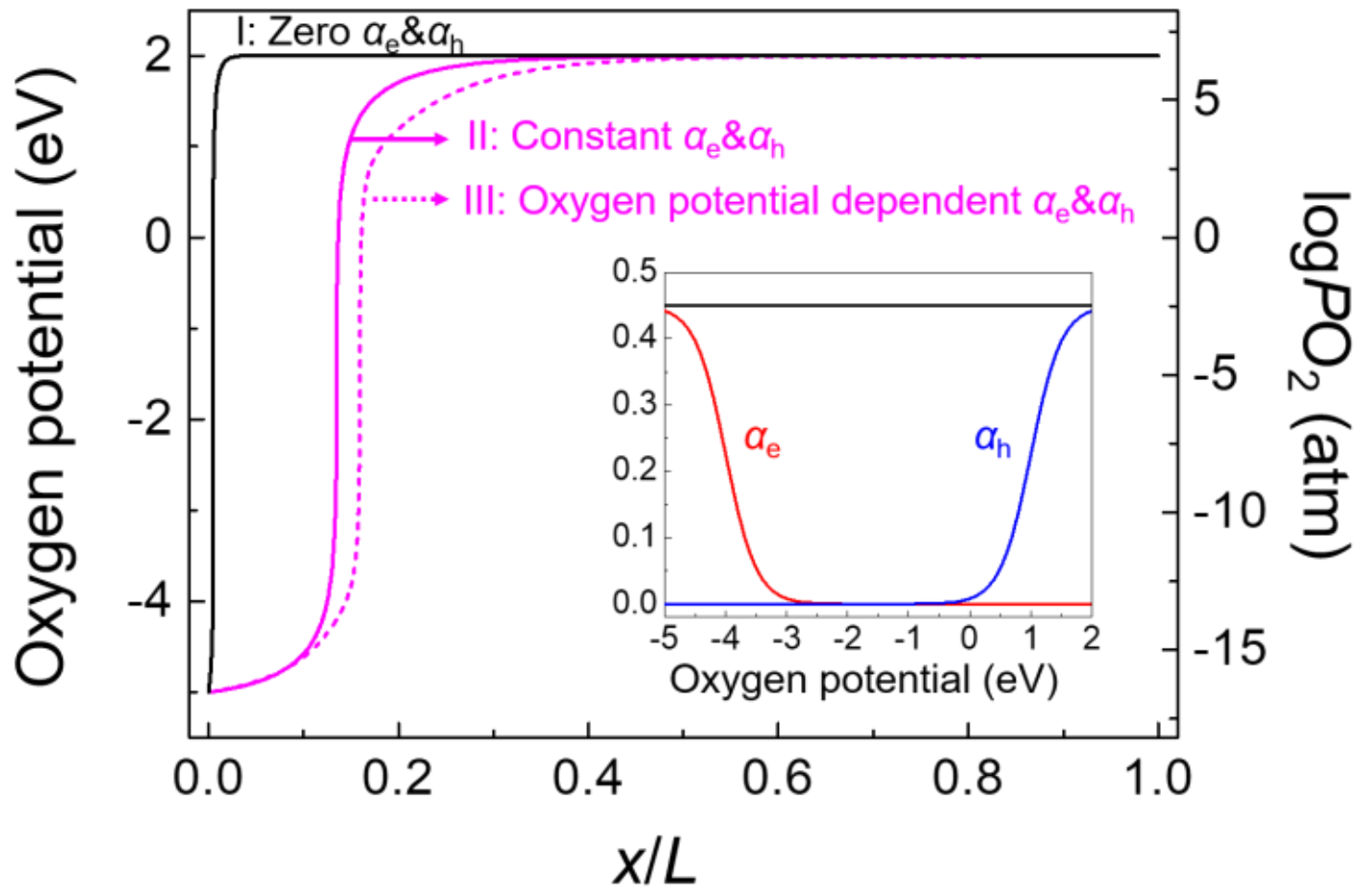

**Figure 8** Calculated spatial distributions of oxygen potential with case I: no internal reaction, $\alpha_e$=$\alpha_h$=0, solid curve in black; II: $\alpha_e$=$\alpha_h$=0.45, solid curve in purple; and III: oxygen-potential-dependent $\alpha_e$ and $\alpha_h$ shown in inset, dash curve in purple. Temperature: 1250 $^o$C, thickness: $L$=1.5 mm, current density: −50 A/cm$^2$, oxygen potential from −5 eV to 2 eV along distance $x$ from left electrode.

(3) p-n junction, electrode polarization, and phase transition

As can be seen from **Fig. 3-4** and **6-7**, a sharp oxygen potential transition always exists, with or without $O^-$ conduction, in both thin and thick electrolytes. This holds as long as the terminal oxygen potentials traverse the two sides of the conductivity minimum in **Fig. 1**. As shown in **Fig. 3a**, this condition is met in a typical SOEC/SOFC whose electrolyte separates the fuel side and the air side. Therefore, the transition will occur in these devices regardless of the loading conditions, electrolyte thickness and electrode polarization, including at temperatures <800 °C and smaller current densities for practical SOFC/SOEC applications. (Grain size transition cannot be observed under these conditions because of negligible grain boundary mobility, but one or two isolated large grains can sometimes be observed next to the severely reduced electrode according to our experience. This should not be mistaken for cathode/anode localization, however, because lacking mobility grain size is no longer a valid marker of oxygen potential.) Our results may have more practical implications. For example, ceria-based electrolyte is known to suffer from large chemical expansions under reduction because $Ce^{3+}$ is larger than $Ce^{4+}$. [41] Given the potential transition that is approximately antisymmetric and lying in the mid-section, the resulting strain profile should also be antisymmetric with a "neutral axis" lying at the mid-section (at the potential of the conductivity minimum in **Fig. 1**), which is the same profile as seen in a bent beam. In fact, fixing the transition near the mid-section will cause a different amount of bending from that caused by a transition near the electrodes, which will in turn lead to a different compensating elastic bending to make the entire section free of bending moment overall. Since it is the latter stress that remains and may possibly result in cracking [18,19], oxygen potential transition and how it depends on the defect charge states may affect device integrity.

As we already proposed in Ref. [16], the transition can be conceptually visualized as an p-n junction: A p-type region with a high hole conductivity, an n-type region with a high electron conductivity, joined by a junction with a huge junction resistance because of minimal electronic conductivity as shown in **Fig. 1**. Without any internal reaction, the electronic current is totally decoupled from the ionic current and cannot receive any assistance from ionic conductivity, so it must face the junction resistance alone despite the fact that the ionic conductivity well exceeds the electronic conductivity everywhere. As the huge junction resistance demands a huge driving force, which is provided by the steep slope of the potential, it gives rise to the potential transition. Using this picture, we can also see that, if the experiment is performed under such condition that the terminal potentials do not traverse the two sides of the conductivity bottom in **Fig. 1**, then there is no p-n junction at all. Indeed, when we performed the experiment in hydrogen gas or argon [16], whose oxygen potential is expected to always lie to the left of the conductivity bottom, we observed a much more gradual variation in grain size without a sharp transition, which corresponds to having the entire electrolyte placed into the n-type region. (The calculated oxygen potential distributions, with $O^-$ conduction, for the above two cases were previously reported as **Fig. 9** in Ref. [16].) These observations provided further support to our analysis of oxygen potential transition.

Lastly, the results in **Fig. 4** and **6** showing an extreme sensitivity of the location of the oxygen potential transition to current density, the balance between electron and hole conductivity, and the conductivity of $O^-$, lead us to foresee conditions that may cause an abrupt propagation of the transition interface. As the sharp transition of oxygen potential is akin to the phase boundary seen in a first-order phase transition, and it is well-known that such interface is prone to instability developing finger-like protrusions and depressions when it propagates under diffusion control, we may envision similar instability with the potential transition. This may be visualized as an abrupt propagation of the p-n junction, followed by its breakdown into a non-planar geometry. Since a high current-density by way of polarization is very effective in altering the boundary potential in addition to being able to sharpen the transition as shown in **Fig. 4**, we expect devices that are subject to high current density may be especially susceptible to an abrupt propagation of the oxygen potential transition front, hence “interface” instability. Such devices span from high temperature electrochemical cells, to multilayer ceramic capacitors, to thin-film resistance memories. Observations in support of this expectation will be presented and analyzed in a forthcoming paper.

## VIII. Conclusions

(1) A general, self-consistent solution to oxygen potential distribution has been obtained in yttria-stabilized zirconia, which allows mixed conduction and internal conversion of $O^{2-}$, $O^{-}$, electrons and holes.

(2) The movement of mixed valence ions such as $O^{-}$ must also involve electron or hole movement. The classical Park and Blumenthal experiment cannot separate the electron and hole conductivity from mixed valence ion conductivity. Instead, it measures the sum of conductivities of electrons, holes and mixed valence ions, which become increasingly important as the temperature lowers. A non-vanishing $O^{-}$ conductivity in turn increases the overall oxygen flow.

(3) The oxygen potential distribution undergoes a sharp step-like transition when the two terminal oxygen potentials lie on the two sides of the characteristic potential where the minimum electronic conductivity lies. When there is adequate grain boundary mobility, such transition would lead a to a grain size transition in YSZ, which has been experimentally observed.

(4) The transition is very sensitive to loading conditions and the balance of electron and hole conductivity. However, $O^{-}$ conductivity has a buffering effect on the transition, making it less sensitive to the variation of loading conditions and electron and hole conductivity. To provide such buffering effect, it is sufficient to have $O^-$ conductivity aided by electrons or holes near the two electrodes.

(5) The sharp oxygen potential transition is akin to a first-order phase transition. Therefore, like in a first-order phase transition, interface instability is anticipated for a moving transition front, which is likely to occur in high current-density devices.

**Acknowledgements**

This work was supported by the Department of Energy (BES grant no. DEFG02-11ER46814) and used the facilities (LRSM) supported by the U.S. National Science Foundation (grant no. DMR-1120901).

**Appendix**

In this appendix, we consider the classical picture of mixed ionic and electronic conduction without internal reactions (i.e. no $O^-$ conduction) and seek to explain why in Park and Blumenthal's experiment [24], the ionic transference number $t_i$ in the pumping YSZ membrane equals ~0.5. The equivalent circuits for the two YSZ members are shown in **Fig. A1a** for the one under OCV condition and in **Fig. A1b** for the pumping one under SOEC condition. For (a)

$$j_{e/h} = -j_{O^{2-}} = \frac{emf}{R_{e/h} + R_{O^{2-}}} \qquad \text{(A1)}$$

For (b)

$$j'_{O^{2-}} = \frac{U - emf}{R_{O^{2-}}} \qquad \text{(A2)}$$

$$j'_{e/h} = \frac{U}{R_{e/h}} \qquad \text{(A3)}$$

At steady state, $j_{O^{2-}} + j'_{O^{2-}} = 0$ implies

$$U = \left( \frac{R_{O^{2-}}}{R_{e/h} + R_{O^{2-}}} + 1 \right) emf \qquad \text{(A4)}$$

In the limit of $R_{e/h} >> R_{O^{2-}}$, we obtain

$$j'_{e/h} \approx j'_{O^{2-}} \approx \frac{emf}{R_{e/h}} \qquad \text{(A5)}$$

$$t_i = \frac{j'_{O^{2-}}}{j'_{e/h} + j'_{O^{2-}}} \approx 0.5 \qquad \text{(A6)}$$

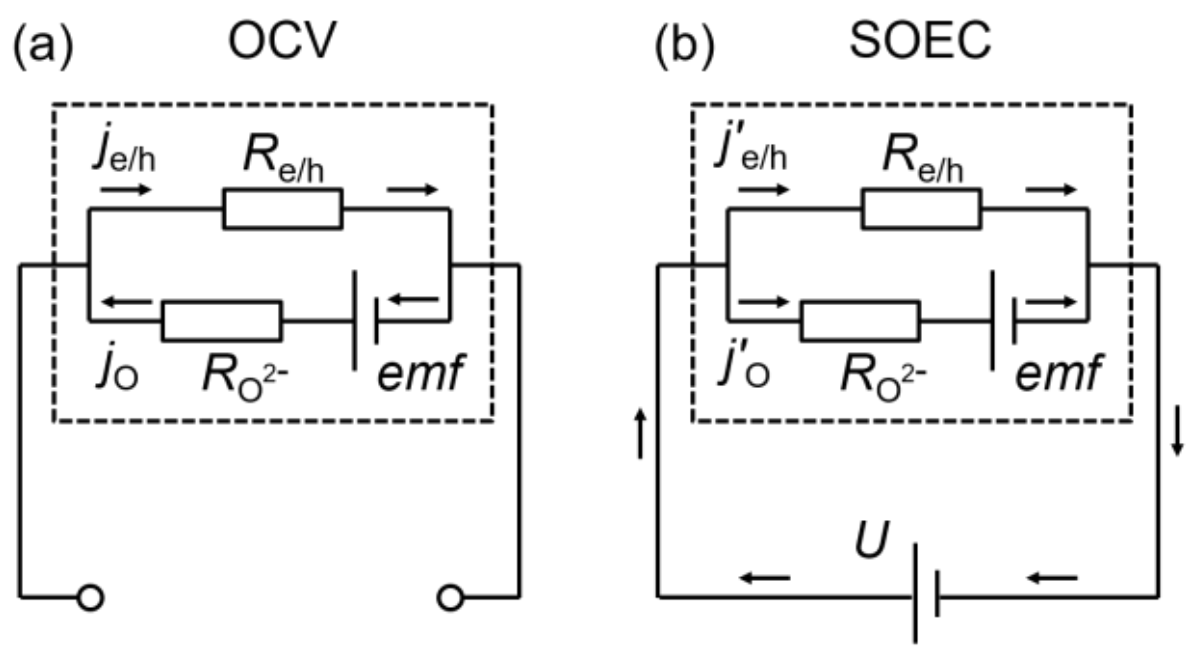

**Figure A1** Equivalent circuits for two YSZ membrane (a) under OCV and (b) SOEC conditions

in Park and Blumenthal's experiment.